\documentclass{refrep}
\usepackage{graphicx}
\usepackage{adjustbox}
\usepackage{rotating}
\usepackage{wasysym}
\newcommand{\simpre}
{
\includegraphics[width=1.2cm]{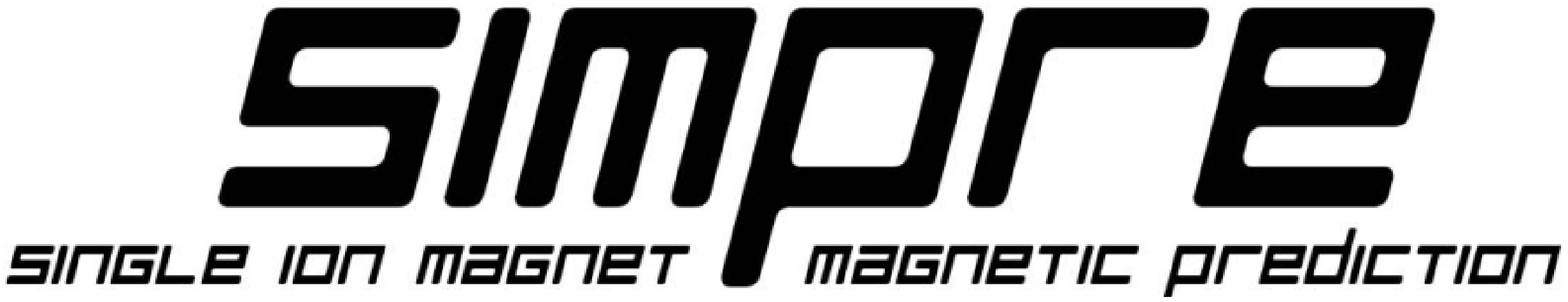}
}
\newcommand{\JaaJe}{\ensuremath{\left(J_++J_-\right)}}
\newcommand{\JaeJe}{\left(J_+-J_-\right)}
\newcommand{\JaaJed}{\left(J_+^2+J_-^2\right)}
\newcommand{\JaeJed}{\left(J_+^2-J_-^2\right)}
\newcommand{\JaaJet}{\left(J_+^3+J_-^3\right)}
\newcommand{\JaeJet}{\left(J_+^3-J_-^3\right)}
\newcommand{\JaaJec}{\left(J_+^4+J_-^4\right)}
\newcommand{\JaeJec}{\left(J_+^4-J_-^4\right)}
\newcommand{\JaaJeq}{\left(J_+^5+J_-^5\right)}
\newcommand{\JaeJeq}{\left(J_+^5-J_-^5\right)}
\newcommand{\JaaJes}{\left(J_+^6+J_-^6\right)}
\newcommand{\JaeJes}{\left(J_+^6-J_-^6\right)}

\begin{document}

\thispagestyle{empty}

\vspace{1.0cm}

\begin{figure*}[h]
\hspace{-1.0cm}
\includegraphics[width=0.8\textwidth]{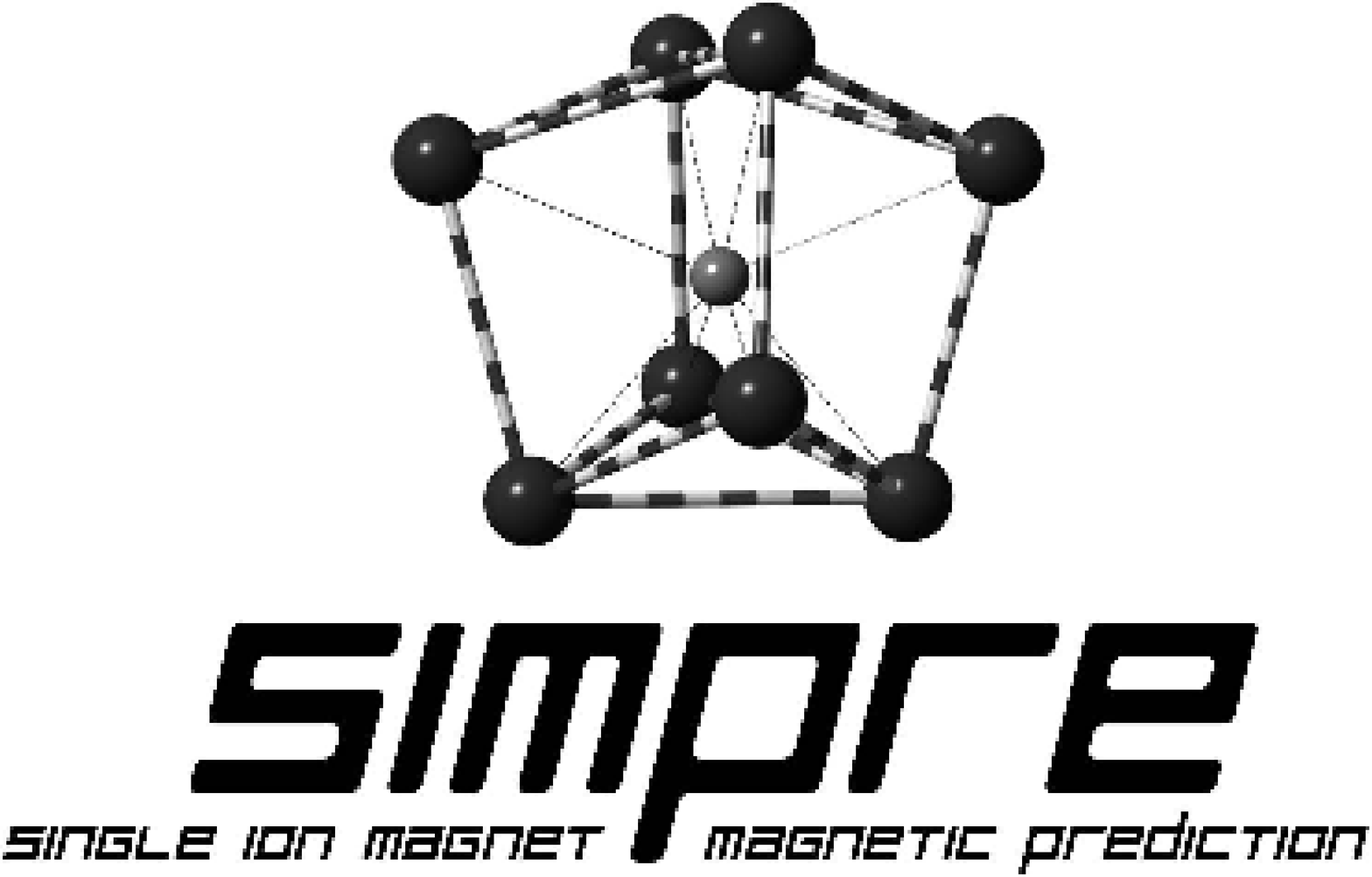}
\end{figure*}

\hspace{-2.0cm}
{\Huge Installation and user manual}

\vspace{3.0cm}


\hspace{1.5cm}
{\Large \today}

\newpage
\tableofcontents

\pagestyle{headings}

\chapter{License}

The \simpre software package is \copyright{} 2013-2014 by Juan M. Clemente-Juan
and Jos\'e J.  Baldov\'{i}.  The present installation and user manual is
\copyright{2013-2014} by Jos\'e J.  Baldov\'{i}, Juan M. Clemente-Juan and
Alejandro Gaita-Ari\~no. The \simpre logo is \copyright{} 2013-2014 by S.
Cardona-Serra.
\\

The source code of \simpre is freely available by request to the authors at
{\verb juan.m.clemente@uv.es } and will be uploaded to the website of the
European Institute of Molecular Magnetism, http://www.eimm.eu/ .

Any published work that contains results obtained using \simpre should cite:
\marginpar{Cite when using \simpre$\Longrightarrow$}Jos\'e J.
Baldov\'{i}, Salvador Cardona-Serra, Juan M. Clemente-Juan, Eugenio Coronado,
Alejandro Gaita-Ari\~no , Andrew Palii, {\em Journal of Computational
Chemistry} 34 (22), 1961-1967, {\bf 2013}. When using improved or patched
versions of this software, and until said versions get published as independent
programs, the same reference should be cited.

To model real systems with \simpre, it is strongly advised to apply the Radial
Effective Charge (REC) model, \marginpar{Use \simpre with
REC$\Longrightarrow$}citing: Jos\'e J.  Baldov\'{i}, Juan J. Borr\'as Almenar,
Salvador Cardona-Serra, Juan M.  Clemente-Juan, Eugenio Coronado, Alejandro
Gaita-Ari\~no, {\em Dalton Transactions Chemistry} 41, 13705, {\bf 2012} \\\\

BECAUSE THE PROGRAM IS LICENSED FREE OF CHARGE, THERE IS NO WARRANTY FOR THE
PROGRAM, TO THE EXTENT PERMITTED BY APPLICABLE LAW. EXCEPT WHEN OTHERWISE
STATED IN WRITING THE PROGRAM IS PROVIDED ``AS IS'' WITHOUT WARRANTY OF ANY KIND,
EITHER EXPRESSED OR IMPLIED, INCLUDING, BUT NOT LIMITED TO, THE IMPLIED
WARRANTIES OF MERCHANTABILITY AND FITNESS FOR A PARTICULAR PURPOSE. 

\chapter{Introduction}

Molecular magnetism applies molecular techniques for designing and studying new
classes of molecule-based magnetic materials, from the bulk to the nanoscale.
Over the past ten years, the field has been revitalized by the discovery that
mononuclear metal complexes may show a single-molecule magnetic (SMM) behavior,
with a new set of unusual quantum physical phenomena as compared with
polynuclear complexes.

In contrast with the classical polynuclear SMMs, whose properties are dominated
by exchange interactions between the ions, in Single Ion Magnets there is a
direct relationship between the electronic spectrum resulting from the crystal
field splitting of the single ion and the magnetic properties of the molecule.
For this reason, there is a need for general models that are capable to
correlate the structural and electronic features of the metal complex with its
SMM properties. 

\simpre is a {\verb fortran77 } code based on an effective electrostatic model
of point charges around a rare earth ion. Although the minimal Point Charge
Electrostatic Model (PCEM) is a very rough approximation, it presents some
interesting features that, together with its beautiful simplicity, makes it
attractive for some purposes. For example, using ideal symmetries it can be
used to reduce the number of adjustable parameters in a fitting procedure.
Moreover, in the case of ionic homoleptic complexes it can be used to predict
the sign of the crystal field parameters. It has been found remarkable that
such a simple model produces good agreement between experimentally and
theoretically determined crystal field parameter signs.$^{\rm 1}$

The program calculates the full set of crystal field parameters, energy levels
and wave functions, as well as the magnetic properties such as magnetization
and the temperature dependence of the magnetic susceptibility.  It is designed
for real systems that need not display ideal symmetry, and is able to
automatically determine the orientation that results in the most compact
description of the wavefunction; which usually -but not always- coincides with
the calculated easy axis of magnetization. 

This handbook includes three examples with full input and output files. When
performing test runs in your own system, please note that technical details
including the compiler version are known to affect the results. Changes in the
order of 1\permil{} in the coefficients of the wavefunctions are
routinely encountered and should not be taken as a sign of a problem.

The systematic application of \simpre to different coordination environments
allows magneto-structural studies. \marginpar{Check out {\bf 4. A step beyond}
for ways of getting more out of \simpre!} The package has already been
successfully applied to several mononuclear systems with single-molecule
magnetic behavior, and, thanks to the parameterization of common ligands as
effective charges, it is possible to build upon these results to not only
rationalize but also predict the properties of more complex systems.

\section{Theoretical model}

This section consists in a complete explanation to be used if a detailed
\emph{Methodology} section is required.

We have resolved a Crystal Field (CF) Hamiltonian where the CF parameters for 
the ground $J$-multiplet have been obtained by a corrected electrostatic point
charge model. For a given $J$-multiplet, such Hamiltonian expressed in terms of
the Extended Stevens Operators (ESOs)$^{\rm 2}$ takes the general form:

\begin{equation}
\widehat{H}_{cf}(J) = \sum_{k=2,4,6}\sum^k_{q=-k}B^q_kO^q_k
= \sum_{k=2,4,6}\sum^k_{q=-k}a_k(1-\sigma_k)A^q_k\langle r^k\rangle O^q_k
\end{equation}

where $k$ (for $f$-elements, $k=2,4,6$) is the order (also called rank or degree)
and $q$ is the operator range (that varies between $k$ and $-k$), of the
Stevens operator equivalents $O^q_k$, as defined by Ryabov in terms of
the angular momentum operators $J_\pm$ and $J_z$ (where the components
$O^q_k$(c) and $O^q_k$(s) correspond to the ESOs with $q\ge0$ and $q<0$
respectively).$^{\rm 3}$ Note that all the Stevens CF parameters $B_k^q$  are
real, whereas the matrix elements of $O^q_k$ ($q < 0$) are imaginary. $a_k$ are
the tabulated $\alpha$, $\beta$ and $\gamma$ Stevens coefficients$^{\rm 4}$ for
$k = 2, 4, 6$, respectively, which are tabulated and depend on the number of
$f$ electrons. $\sigma_k$ are the Sternheimer shielding parameters$^{\rm 5}$ of
the 4f electronic shell and $\langle r^k\rangle$ are the expectation values of
the radius.$^{\rm 5}$

In \simpre 1.1,${\rm ^6}$ the $A_k^q$ parameters are determined by the following relations:
\begin{equation}
A_k^0 = \frac{4\pi}{2k+1}\sum_{i=1}^N \frac{Z_i e^2}{R_i^{k+1}}Z_{k0}\left(\theta_i,\phi_i\right)p_{kq} \qquad\qquad
\end{equation}
\begin{equation}
A_k^q = \frac{4\pi}{2k+1}\sum_{i=1}^N \frac{Z_i e^2}{R_i^{k+1}}Z^c_{kq}\left(\theta_i,\phi_i\right)p_{kq} ; (q>0) 
\end{equation}
\begin{equation}
A_k^q = \frac{4\pi}{2k+1}\sum_{i=1}^N \frac{Z_i e^2}{R_i^{k+1}}Z^s_{k|q|}\left(\theta_i,\phi_i\right)p_{k|q|} ; (q<0) 
\end{equation}
where $R_i$,$\theta_i$,$\phi_i$, are the effective polar coordinates of the
point charges with the lanthanoid at the origin; $Z_i$ is the effective point
charge, associated with the $i$-th donor atom; $N$ is the number of effective
charges; $e$ is the electron charge; $p_{kq}$ are the prefactors of the
spherical harmonics and $Z_{kq}$ are the tesseral harmonics expressed in terms
of the polar coordinates for the $i$-th donor atom. 

\newpage
{\bf References}

{\rm [1]} C. G\"orller-Walrand, K. Binnemans, Rationalization of Crystal-Field
Para\-me\-trization, \emph{Handbook of Physics and Chemistry of Rare Earths}, vol.
23, eds. K. A. Gschneidner, L. Eyring, 1996, Elsevier

{\rm [2]} a) C. Rudowicz, C.Y. Chung, \emph{J. Phys. Condens. Matter} 2004,
{\bf 16}, 5825; b) C. Rudowicz, {\emph J. Phys. C: Solid State Phys.} 1985,
{\bf 18}, 1415; c) C.  Rudowicz, {\emph J. Phys. C: Solid State Phys.} 1985b
{\bf 18}, 3837 (erratum).

{\rm [3]} I. D. Ryabov, {\emph Journal of Magnetic Resonance} 1999, {\bf 140},
141

{\rm [4]} K. W. H. Stevens, {\emph Proc. Phys. Soc.} 1952, {\bf 65}, 209

{\rm [5]} S. Edvardsson, M. Klintenberg, {\emph J. Allow. Comp.} 1998, {\bf
230}, 275

{\rm [6]} J. J. Baldov\'i, J. M. Clemente-Juan, E. Coronado, A. Gaita-Ari\~no, A. Palii,
{\emph J. Comput. Chem.}, 2014 DOI: 10.1002/jcc.23699

\chapter{User guide}

\section{System requirements}
\simpre is not distributed as operating-system-specific binary but as a
portable {\verb fortran77 } code. Thus, its use requires a standard 
{\verb fortran77 } compiler (contact your local system administrator for help if
needed).  So far it has been tested with {\verb gfortran } ,
{\verb ifort } and {\verb g77 }, and it has been optimized for {\verb gfortran }.
Additionally, it requires the {\verb lapack } library, which can be freely
obtained from: \marginpar{Remember, \simpre will {\bf not} compile without
lapack! } http://www.netlib.org/lapack/ .

\simpre is a lightweight program, which will run even on a linux netbook or an
old desktop computer (1 GB RAM, 1.0 GHz processor). With the rotation option
turned off ({\verb irot=0 }, see below), execution is very fast and in practice
the limiting step is compilation, which takes a few seconds. Compiling is
required for every change in the input files and/or when moving to a different
environment, e.g. to a different computer or operative system. Execution with
full rotation can take about an hour in a minimal system and just a few minutes
on a more powerful workstation. As described in the next section, you will also
need a plain text editor -such as vi, TextEdit or Notepad- to prepare the input
files.

\section{Input files and syntax}

The input of \simpre is divided into two files: 
\begin{itemize}
\item {\verb simpre.par }, which defines computational parameters and yes/no
switches for the calculation and
\item{\verb simpre.dat }, which defines the actual data: the metal and the
point charges, but also, if needed, the output options for the susceptibility
and the magnetization.
\end{itemize}

\subsection{simpre.par}

As {\verb simpre.par } is taken as an {\verb include } by {\verb simpre.f }, it
is a file in standard {\verb fortran 77 } format, i.e.: \marginpar{\simpre will
{\bf not} compile if simpre.par is not valid fortran77 !}
the text is case insensitive, you need to leave 6 blank characters at the
beginning of each line containing a command such as  {\verb parameter }, every
line beginning by {\verb C } is a comment and the rest of the line after a
{\verb ! } is also a comment. It defines nine parameters: 
\begin{itemize}
\item {\verb idtot } is the dimension of the full energy matrix, which is
calculated as $idtot=2J+1$. For $f$ elements it is safe to leave it as the
maximum possible value $17=2\cdot8+1$.
\item {\verb iuni } defines the energy (or energy equivalent) units for the
output: 1 for equivalent temperature in K, 2 for energy in meV, 3 for
wavenumber in cm$^{-1}$ 
\item {\verb icoo } is the type of coordinates in {\verb simpre.dat }: 1
for spherical (radius $r$ in \AA, polar angle $\theta$ and azimuthal angle
$\phi$ in $^\circ$) , 2 for cartesian ($x, y, z$ in \AA)
\item {\verb ishi }, a yes/no (1/0) switch, controls the use of Sternheimer
shielding parameters
\item {\verb irot } (1/0) switch controls whether a rotation of the
coordinates searching for the most compact description of the wavefunction has
to be performed or not
\item the {\verb isus } (1/0) switch controls whether the magnetic
susceptibility will be calculated
\item the {\verb imag } (1/0) switch controls whether the magnetization
will be calculated
\item {\verb ieig } decides whether the eigenvalues are going to be included in
the output (1) or not (0), or if not only the eigenvalues but also the
eigenvectors will be included (2)
\item {\verb EPS } is a numerical parameter to deal with underflow problems, it
is safe to leave it as $10^{-12}$ ({\verb 1.D-12 }) and not change it unless
dealing with technical problems in a particular computer
\item {\verb ICHAR } 
\marginpar{Remember to adjust the number of charges as needed!}
is the number of charges to be accepted as input
in {\verb simpre.dat }; this is needed for forward-compatibility reasons, as
future versions of \simpre may deal with polynuclear complexes. If a high
number of charges needs to be used for any reason, {\verb ICHAR } should be
changed accordingly.
\end{itemize}

\subsection{simpre.dat}

{\verb simpre.dat } needs to define the metal with the following code, which is
included as first line of the input as a reminder:
\maxipagerulefalse
\begin{maxipage}
\begin{tabular}{c|c|c|c|c|c|c|c|c|c|c|c}
  1  &  2  &  3  &  4  &  5  &  6  &  7  &  8  &  9  &  10  &  11  &  12  \\
\hline
Ce$^{3+}$&Pr$^{3+}$&Nd$^{3+}$&Pm$^{3+}$&Sm$^{3+}$&Tb$^{3+}$&Dy$^{3+}$&Ho$^{3+}$&Er$^{3+}$&Tm$^{3+}$&Yb$^{3+}$&user-defined\\
\end{tabular}
\end{maxipage}

Thus, after the reminder and a blank line, the third line starts with an
integer code (length 3) from 1 to 12 that defines the metal ion. If the user
chooses the code {\verb 12 }, the lower part of the file defines the metal, as
we will see below. 

In the fourth line, the user defines the number of effective charges, again
as a length 3 integer. This is followed by a list containing, in five columns,
the ordinal of each charge (length 3 integer), the coordinates (cartesian or
spherical, as specified in {\verb simpre.par }) and the magnitude of each
charge. Both the coordinates and the charge are length 13, with the decimal
point in position 6. \marginpar{Do not use negative numbers for the charge.}
Note that the charge in the input is a {\bf positive} number which is
understood as the fraction of a (negative) electron charge.  The coordinates
are in \AA{} and $^\circ$.

Immediately after the charge list, two extra lines detail the desired output
for the magnetization and susceptibility.  If {\verb simpre.par } indicates
that magnetization and/or susceptibility need to be calculated, they will be
read.  \marginpar{These lines should not be deleted.} Otherwise, they will be
ignored.  Nevertheless, the lines themselves need to be there.  For the
susceptibility, the input indicates the minimum and maximum temperatures (in K,
length 9 with the decimal point in position 6), the number of points (length 6
integer) and the applied magnetic field (in mT, length 9 with the decimal point
in position 7).  For the magnetization, the input indicates the temperature (in
K, length 9 with the decimal point in position 6), the maximum field (mT,
length 9 integer), the field increment (mT, length 6 integer) and a number
defining the angular integration (length 9 integer). Large numbers (100 would
be large in this context) for angular integration mean longer calculations but
produce more precise values for average magnetization and average $\chi T$
products, specially in very anisotropic environments. After this two lines
comes a mandatory blank line, meaning {\bf at least three lines need to be
present after the coordinate list}. 

You only need to care about the last part of the input if there is a
user-defined metal.  The first line indicates the symbol for the metal, its
angular momentum as $2J$ and the value of $g_J$. The second line indicates the
Stevens parameters $\alpha, \beta, \gamma$. The third line indicates the
shielding parameters $\sigma_2, \sigma_4, \sigma_6$. The last line defines the
expectation values of the radius powers $\langle r^2\rangle, \langle
r^4\rangle, \langle r^6\rangle$. This is also a way of changing the default
choice for $\langle r^k\rangle$\footnote{Compare Freeman and Watson,
\emph{Phys. Rev.} 127, 2058 (1962) with Freeman and Desclaux, \emph{J. Magn.
Magn.  Mater.} 12, 11 (1979)} (and/or for the shielding parameters).

\section{Output files and interpretation}

\subsection{simpre.out}

At the very top of {\verb simpre.out } one can see a standard message including
the version number and the program name in ASCII art. 

The first real information is a repetition of some of the input data. We can
read the name of the ion, the number of charges, the value of the Land\'e $g$
factor for said ion and the energy units, followed by a small table for the
Stevens parameters $\alpha$, $\beta$, $\gamma$, the radius expectation values
$\langle r^2 \rangle$, $\langle r^4 \rangle$, $\langle r^6 \rangle$ and the
shielding parameters $\sigma^2$, $\sigma^4$ and $\sigma^6$ (including the
contribution from the 4f perturbation), if {\verb ishi = 1}.

Next come the input coordinates, which are given both as cartesian and as
spherical, with distances in \AA{} and angles in radians. If the rotation
option was activated, this is followed by the applied Euler angles and the
spherical coordinates after rotation that produce the most compact description
of the wavefunction, i.e. the orientation that allows grouping the highest
weight onto a single $M_J$ component in the ground state.  \marginpar{Check out
{\bf 4. A step beyond} for ways of getting more out of \simpre!} This does not
always coincide with the easy axis of magnetization, and indeed in some
complexes it can be precisely the hard axis of magnetization. In any case, it
is always possible to obtain the easy axis by rotating the input coordinates
manually and calculating the parallel magnetization until reaching the maximum.
Improved versions of \simpre exist that can automatize this procedure.

These coordinates define the reference frame for the $A_k^q$ Stevens
parameters, which are given as a table in the form $A_k^q\langle r^k\rangle$ and
$B_k^q$.

Finally, if requested, the list of eigenvalues and eigenvectors are included,
using all $M_J$ from the ground $J$ multiplet as basis set. First the
coefficients are written as complex numbers (first line: real part, second
line: imaginary part), and later as the squares of the moduli (i.e.: normalized
to unity) to facilitate the comparison of the different contributions. Note
that energies are given with a precision up to the millionth of the unit to
allow for testing, but for most purposes it makes more sense to round to the
nearest unit.

\subsection{sus.out}

At the very top of {\verb simpre.out } one can see a standard message including
the version number and the program name in ASCII art. 

Then the relevant input data are repeated: minimum and maximum temperature,
number of points and value of the magnetic field.

Finally, the result is given in five columns: temperature in K, and $\chi T$
product in emu$\cdot$K/mol for the different orientations and the average:
$\chi_z T$, $\chi_x T$, $\chi_y T$, $\chi_{av} T$.

\subsection{mag.out}

At the very top of {\verb simpre.out } one can see a standard message including
the version number and the program name in ASCII art. 

Then the relevant input data are repeated: temperature, maximum field and field
step and integration angle. The integration angle is given indirectly in the
input as number of points.

Finally, the result is given in five columns: field in T and magnetization in
Bohr Magnetons for the different orientations and the average: $M_z$,
$M_x$, $M_y$, $M_{av}$.

\newpage
\section{Example 1: Ideal cube (DyX$_8$)}

This is an idealized example that can serve as a test run, or it could also be
part of a magnetostructural study. The center ion is Dy$^{3+}$ and the
environment consists in 8 unity charges in the corners of a cube, each at a
distance of 2.5\AA{} from the center. There are three mandatory lines (blank,
in this case) after the charge list.

\subsection*{simpre.par}

\begin{small}
\begin{maxipage}
\begin{verbatim}
      PARAMETER (idtot= 17)! dimension of full energy matrix
c
      PARAMETER (iuni=3)  ! units (1=K, 2=meV, 3=cm-1)
c
      PARAMETER (icoo=1)  ! spherical coordinates=1, cartesian coordinates=2
c
      PARAMETER (ishi=1)  ! use of Sternheimer shielding parameters [1(yes)/0(no)]
c
      PARAMETER (irot=0)  ! rotation for a compact ground-state wave function [1(yes)/0(no)]
c
      PARAMETER (isus=0)  ! magnetic susceptibility calculation [1(yes)/0(no)]
c
      PARAMETER (imag=0)  ! magnetization calculation [1(yes)/0(no)]
c
      PARAMETER (ieig=2)  ! write eigenvalues 
c                                   [no(0)/yes(1)/yes+eigenvectors(2)]
c
c------------------------------------------------------------------------------
c* The following parameter is defined to minimize underflow problems
        PARAMETER (EPS    = 1.D-12)
c
        PARAMETER (ICHAR =  12)! number of charges
\end{verbatim}
\end{maxipage}
\end{small}

\subsection*{simpre.dat}

\begin{small}
\begin{maxipage}
\begin{verbatim}
1=Ce3+, 2=Pr3+, 3=Nd3+, 4=Pm3+, 5=Sm3+, 6=Tb3+, 7=Dy3+, 8=Ho3+, 9=Er3+, 10=Tm3+, 11=Yb3+, 12=user

  7    !ion code (from 1 to 12, see above)
  8    !number of effective charges
  1    2.5000000   54.7356103    0.0000000    1.0000000
  2    2.5000000   54.7356103   90.0000000    1.0000000
  3    2.5000000   54.7356103  180.0000000    1.0000000
  4    2.5000000   54.7356103  270.0000000    1.0000000
  5    2.5000000  125.2643897    0.0000000    1.0000000
  6    2.5000000  125.2643897   90.0000000    1.0000000
  7    2.5000000  125.2643897  180.0000000    1.0000000
  8    2.5000000  125.2643897  270.0000000    1.0000000



\end{verbatim}
\end{maxipage}
\end{small}

\newpage
\subsection*{simpre.out}

\begin{scriptsize}
\hspace{-6.0cm}\begin{maxipage}
\begin{verbatim}
******************************************************
**  output file generated by SIMPRE version 1.1  **
******************************************************

 ____________________________________________________

 .######..####.##.....##.########..########..########
 ##....##..##..###...###.##.....##.##.....##.##......
 ##........##..####.####.##.....##.##.....##.##......
 .######...##..##.###.##.########..########..######..
 ......##..##..##.....##.##........##...##...##......
 ##....##..##..##.....##.##........##....##..##......
 .######..####.##.....##.##........##.....##.########
 ____________________________________________________

 ****************************************************

               ion:      Dy(III)
 number of charges:            8
                 g:        1.333
             units:         cm-1

 stv           2           4           6
     -0.00634921 -0.00005920  0.00000103

 <r>k          2           4           6
      0.84900000  1.97700000 10.44000000

 shi           2           4           6
      0.52700000 -0.01990000 -0.03160000


  input coordinates

  cartesian (A)
         x            y            z            Z  
    2.0412415    0.0000000    1.4433757    1.0000000
    0.0000000    2.0412415    1.4433757    1.0000000
   -2.0412415    0.0000000    1.4433757    1.0000000
   -0.0000000   -2.0412415    1.4433757    1.0000000
    2.0412415    0.0000000   -1.4433757    1.0000000
    0.0000000    2.0412415   -1.4433757    1.0000000
   -2.0412415    0.0000000   -1.4433757    1.0000000
   -0.0000000   -2.0412415   -1.4433757    1.0000000

  spherical (rad)
         r          theta         phi           Z  
    2.5000000    0.9553166    0.0000000    1.0000000
    2.5000000    0.9553166    1.5707963    1.0000000
    2.5000000    0.9553166    3.1415927    1.0000000
    2.5000000    0.9553166    4.7123890    1.0000000
    2.5000000    2.1862760    0.0000000    1.0000000
    2.5000000    2.1862760    1.5707963    1.0000000
    2.5000000    2.1862760    3.1415927    1.0000000
    2.5000000    2.1862760    4.7123890    1.0000000

 **************************************************** 
     Stevens Crystal Field Parameters (cm-1)          
 **************************************************** 

   k   q         Akq <rk>             Bkq             
 ---------------------------------------------------- 

   2   0        0.00000142       -0.00000001          
   4   0      -73.12735068        0.00432914          
   4   4      365.63675289       -0.02164572          
   4  -4        0.00000000        0.00000000          
   6   0        5.00013438        0.00000517          
   6   4      105.00282221        0.00010867       
   6  -4        0.00000000        0.00000000       
  \end{verbatim}
  \end{maxipage}
\end{scriptsize}

\newpage

\begin{scriptsize}
\hspace{-5.0cm}\adjustbox{minipage=\textwidth,angle=-90}{%
\begin{verbatim}
Eigenvalues (cm-1) and eigenvectors

                     -7.50     -6.50     -5.50     -4.50     -3.50     -2.50     -1.50     -0.50      0.50      1.50      2.50      3.50      4.50      5.50      6.50      7.50

        0.000000  0.000000  0.000000  0.941752 -0.000000 -0.000000  0.000000  0.271189 -0.000000 -0.000000  0.000000  0.197992 -0.000000 -0.000000  0.000000  0.018963  0.000000
                 -0.000000  0.000000  0.000000 -0.000000  0.000000  0.000000  0.000000 -0.000000  0.000000  0.000000  0.000000 -0.000000  0.000000  0.000000  0.000000  0.000000

        0.000000  0.000000  0.018963 -0.000000 -0.000000 -0.000000  0.197992 -0.000000 -0.000000 -0.000000  0.271189 -0.000000 -0.000000 -0.000000  0.941752 -0.000000  0.000000
                  0.000000  0.000000 -0.000000 -0.000000 -0.000000  0.000000 -0.000000 -0.000000 -0.000000  0.000000 -0.000000 -0.000000 -0.000000  0.000000 -0.000000  0.000000

        0.000002 -0.000000  0.000000 -0.000000  0.798851 -0.000000  0.000000 -0.000000  0.419831 -0.000000  0.000000  0.000000  0.430642 -0.000000  0.000000 -0.000000 -0.011195
                 -0.000000  0.000000  0.000000  0.000000  0.000000  0.000000  0.000000  0.000000 -0.000000  0.000000  0.000000  0.000000  0.000000  0.000000  0.000000  0.000000

        0.000002 -0.011195  0.000000  0.000000  0.000000  0.430642  0.000000  0.000000  0.000000  0.419831  0.000000  0.000000  0.000000  0.798851  0.000000  0.000000 -0.000000
                  0.000000 -0.000000  0.000000  0.000000 -0.000001 -0.000000  0.000000  0.000000 -0.000001 -0.000000  0.000000  0.000000 -0.000001 -0.000000  0.000000  0.000000

       30.943069  0.000000  0.773482 -0.000000  0.000000 -0.000000  0.485988  0.000000 -0.000000 -0.000000  0.344246  0.000000 -0.000000  0.000000 -0.216878  0.000001  0.000000
                  0.000000  0.000000  0.000000  0.000000 -0.000000  0.000000 -0.000000 -0.000000 -0.000000  0.000000 -0.000000 -0.000000  0.000000 -0.000000 -0.000000  0.000000

       30.943069  0.000000 -0.000001 -0.216878 -0.000000 -0.000000 -0.000000  0.344246  0.000000 -0.000000 -0.000000  0.485988  0.000000  0.000000  0.000000  0.773482 -0.000000
                  0.000000  0.000000  0.000000 -0.000000 -0.000000  0.000000 -0.000000  0.000000 -0.000000  0.000000 -0.000000  0.000000  0.000000 -0.000000 -0.000000  0.000000

       30.943070  0.026523 -0.000000  0.000000 -0.000000 -0.795615 -0.000000 -0.000000  0.000000 -0.240210 -0.000000 -0.000000  0.000000  0.555510  0.000000 -0.000000 -0.000000
                 -0.000000  0.000000 -0.000000 -0.000000  0.000000  0.000000  0.000000  0.000000  0.000000  0.000000  0.000000  0.000000 -0.000000 -0.000000  0.000000  0.000000

       30.943070  0.000000 -0.000000 -0.000000  0.555510 -0.000000 -0.000000  0.000000 -0.240210 -0.000000 -0.000000  0.000000 -0.795615  0.000000  0.000000  0.000000  0.026523
                  0.000000 -0.000000 -0.000000  0.000000 -0.000000 -0.000000  0.000000 -0.000000  0.000000 -0.000000  0.000000 -0.000000 -0.000000  0.000000  0.000000  0.000000

       35.821651  0.000000 -0.633278 -0.000000 -0.000000 -0.000000  0.581843  0.000001  0.000000 -0.000000  0.450694  0.000001  0.000000  0.000000 -0.239357 -0.000001 -0.000000
                  0.000000 -0.000000 -0.000000 -0.000000 -0.000000  0.000000  0.000000  0.000000  0.000000  0.000000  0.000000  0.000000 -0.000000 -0.000000 -0.000000 -0.000000

       35.821651 -0.000000 -0.000001  0.239357  0.000000  0.000000  0.000001 -0.450694  0.000000  0.000000  0.000001 -0.581843  0.000000 -0.000000 -0.000000  0.633278 -0.000000
                 -0.000000 -0.000000  0.000000 -0.000000 -0.000000  0.000000 -0.000000  0.000000 -0.000000  0.000000 -0.000000  0.000000  0.000000 -0.000000  0.000000  0.000000

      138.450427 -0.549062  0.000000  0.000000  0.000000  0.312086  0.000000  0.000000 -0.000000 -0.677614  0.000000 -0.000000 -0.000000  0.180183 -0.000000  0.000000  0.000000
                 -0.192542  0.000000  0.000000  0.000000  0.109441  0.000000 -0.000000 -0.000000 -0.237622  0.000000  0.000000 -0.000000  0.063186  0.000000  0.000000  0.000000

      138.450427  0.000000  0.000000  0.000000 -0.190941  0.000000 -0.000000 -0.000000  0.718070 -0.000000  0.000000  0.000000 -0.330719  0.000000 -0.000000 -0.000000  0.581844
                  0.000000  0.000000 -0.000000 -0.000000  0.000000 -0.000000  0.000000  0.000000 -0.000000  0.000000 -0.000000 -0.000000  0.000000 -0.000000  0.000000  0.000000

      141.734248 -0.812791 -0.000000 -0.000000 -0.000000 -0.268642  0.000000  0.000000 -0.000000  0.500416 -0.000000 -0.000000 -0.000000 -0.129562  0.000000  0.000000 -0.000000
                 -0.000017 -0.000000  0.000000 -0.000000 -0.000006  0.000000 -0.000000  0.000000  0.000010 -0.000000  0.000000 -0.000000 -0.000003  0.000000 -0.000000  0.000000

      141.734248 -0.000000 -0.000000 -0.000000  0.129562 -0.000000  0.000000  0.000000 -0.500416  0.000000 -0.000000 -0.000000  0.268642 -0.000000  0.000000  0.000000  0.812791
                 -0.000000 -0.000000  0.000000  0.000000 -0.000000  0.000000 -0.000000 -0.000000  0.000000 -0.000000  0.000000  0.000000 -0.000000  0.000000 -0.000000  0.000000

      141.734252  0.000000 -0.018007  0.000019  0.000000  0.000000  0.621348 -0.000161  0.000000 -0.000000 -0.777706  0.000128 -0.000000  0.000000  0.093682 -0.000004  0.000000
                 -0.000000 -0.000000 -0.000000  0.000000 -0.000000  0.000000  0.000000  0.000000  0.000000 -0.000000 -0.000000 -0.000000 -0.000000  0.000000  0.000000  0.000000

      141.734252  0.000000 -0.000004 -0.093682 -0.000000  0.000000  0.000128  0.777706 -0.000000 -0.000000 -0.000161 -0.621348 -0.000000  0.000000  0.000019  0.018007  0.000000
                  0.000000  0.000000  0.000000 -0.000000  0.000000 -0.000000 -0.000000  0.000000 -0.000000  0.000000  0.000000 -0.000000  0.000000 -0.000000 -0.000000  0.000000
\end{verbatim}
}
\end{scriptsize}

\newpage

\begin{scriptsize}
\hspace{-3.0cm}\adjustbox{minipage=\textwidth,angle=-90}{%
\begin{verbatim}
Eigenvalues (cm-1) and eigenvectors (modulus square)

                     -7.50     -6.50     -5.50     -4.50     -3.50     -2.50     -1.50     -0.50      0.50      1.50      2.50      3.50      4.50      5.50      6.50      7.50

        0.000000  0.000000  0.000000  0.886896  0.000000  0.000000  0.000000  0.073544  0.000000  0.000000  0.000000  0.039201  0.000000  0.000000  0.000000  0.000360  0.000000

        0.000000  0.000000  0.000360  0.000000  0.000000  0.000000  0.039201  0.000000  0.000000  0.000000  0.073544  0.000000  0.000000  0.000000  0.886896  0.000000  0.000000

        0.000002  0.000000  0.000000  0.000000  0.638163  0.000000  0.000000  0.000000  0.176258  0.000000  0.000000  0.000000  0.185453  0.000000  0.000000  0.000000  0.000125

        0.000002  0.000125  0.000000  0.000000  0.000000  0.185453  0.000000  0.000000  0.000000  0.176258  0.000000  0.000000  0.000000  0.638163  0.000000  0.000000  0.000000

       30.943069  0.000000  0.598275  0.000000  0.000000  0.000000  0.236184  0.000000  0.000000  0.000000  0.118505  0.000000  0.000000  0.000000  0.047036  0.000000  0.000000

       30.943069  0.000000  0.000000  0.047036  0.000000  0.000000  0.000000  0.118505  0.000000  0.000000  0.000000  0.236184  0.000000  0.000000  0.000000  0.598275  0.000000

       30.943070  0.000703  0.000000  0.000000  0.000000  0.633004  0.000000  0.000000  0.000000  0.057701  0.000000  0.000000  0.000000  0.308592  0.000000  0.000000  0.000000

       30.943070  0.000000  0.000000  0.000000  0.308592  0.000000  0.000000  0.000000  0.057701  0.000000  0.000000  0.000000  0.633004  0.000000  0.000000  0.000000  0.000703

       35.821651  0.000000  0.401042  0.000000  0.000000  0.000000  0.338542  0.000000  0.000000  0.000000  0.203125  0.000000  0.000000  0.000000  0.057292  0.000000  0.000000

       35.821651  0.000000  0.000000  0.057292  0.000000  0.000000  0.000000  0.203125  0.000000  0.000000  0.000000  0.338542  0.000000  0.000000  0.000000  0.401042  0.000000

      138.450427  0.338542  0.000000  0.000000  0.000000  0.109375  0.000000  0.000000  0.000000  0.515625  0.000000  0.000000  0.000000  0.036458  0.000000  0.000000  0.000000

      138.450427  0.000000  0.000000  0.000000  0.036458  0.000000  0.000000  0.000000  0.515625  0.000000  0.000000  0.000000  0.109375  0.000000  0.000000  0.000000  0.338542

      141.734248  0.660629  0.000000  0.000000  0.000000  0.072168  0.000000  0.000000  0.000000  0.250416  0.000000  0.000000  0.000000  0.016786  0.000000  0.000000  0.000000

      141.734248  0.000000  0.000000  0.000000  0.016786  0.000000  0.000000  0.000000  0.250416  0.000000  0.000000  0.000000  0.072168  0.000000  0.000000  0.000000  0.660629

      141.734252  0.000000  0.000324  0.000000  0.000000  0.000000  0.386073  0.000000  0.000000  0.000000  0.604826  0.000000  0.000000  0.000000  0.008776  0.000000  0.000000

      141.734252  0.000000  0.000000  0.008776  0.000000  0.000000  0.000000  0.604826  0.000000  0.000000  0.000000  0.386073  0.000000  0.000000  0.000000  0.000324  0.000000
\end{verbatim}
}
\end{scriptsize}

\newpage
\section{Example 2: LiHo$_x$Y$_{(1-x)}$F$_4$}

This is a well-known solid-state example. In this case, we simplified the input
by taking average radiuses. As susceptibility and magnetization are requested
in {\verb simpre.par }, the corresponding details are introduced in 
{\verb simpre.dat }. There is a mandatory (blank) line after the details for
the magnetization.

\subsection*{simpre.par}

\begin{small}
\begin{maxipage}
\begin{verbatim}
      PARAMETER (idtot= 17)! dimension of full energy matrix
c
      PARAMETER (iuni=3)  ! units (1=K, 2=meV, 3=cm-1)
c
      PARAMETER (icoo=1)  ! spherical coordinates=1, cartesian coordinates=2
c
      PARAMETER (ishi=1)  ! use of Sternheimer shielding parameters [1(yes)/0(no)]
c
      PARAMETER (irot=1)  ! rotation for a compact ground-state wave function [1(yes)/0(no)]
c
      PARAMETER (isus=1)  ! magnetic susceptibility calculation [1(yes)/0(no)]
c
      PARAMETER (imag=1)  ! magnetization calculation [1(yes)/0(no)]
c
      PARAMETER (ieig=2)  ! write eigenvalues 
c                                   [no(0)/yes(1)/yes+eigenvectors(2)]
c
c------------------------------------------------------------------------------
c* The following parameter is defined to minimize underflow problems
        PARAMETER (EPS    = 1.D-12)
c
        PARAMETER (ICHAR =  12)! number of charges
\end{verbatim}
\end{maxipage}
\end{small}

\subsection*{simpre.dat}

\begin{small}
\begin{maxipage}
\begin{verbatim}
1=Ce3+, 2=Pr3+, 3=Nd3+, 4=Pm3+, 5=Sm3+, 6=Tb3+, 7=Dy3+, 8=Ho3+, 9=Er3+, 10=Tm3+, 11=Yb3+, 12=user

  8    !ion code (from 1 to 12, see above)
  8    !number of centers coordinated to the lanthanoid
  1    1.4893000   67.0800000    0.0000000    0.2003000
  2    1.4893000  112.9200000   90.0000000    0.2003000
  3    1.4893000   67.0800000  180.0000000    0.2003000
  4    1.4893000  112.9200000  270.0000000    0.2003000
  5    1.4893000  142.0500000    3.9800000    0.2003000
  6    1.4893000   37.9500000   93.9800000    0.2003000
  7    1.4893000  142.0500000  183.9800000    0.2003000
  8    1.4893000   37.9500000  273.9800000    0.2003000
    2.000  300.000    25   100.00   !SUSCEPTIBILITY: Tmin(K),Tmax(K),# of points,H(mT) 
    2.000     5000   250       61   !MAGNETIZATION:  T(K),Hmax(mT),H increment (mT),# of angles	
\end{verbatim}
\end{maxipage}
\end{small}

\newpage

\subsection*{simpre.out}

\begin{footnotesize}
\hspace{-6.0cm}\begin{maxipage}
\begin{verbatim}
******************************************************
**  output file generated by SIMPRE version 1.1  **
******************************************************

 ____________________________________________________

 .######..####.##.....##.########..########..########
 ##....##..##..###...###.##.....##.##.....##.##......
 ##........##..####.####.##.....##.##.....##.##......
 .######...##..##.###.##.########..########..######..
 ......##..##..##.....##.##........##...##...##......
 ##....##..##..##.....##.##........##....##..##......
 .######..####.##.....##.##........##.....##.########
 ____________________________________________________

 ****************************************************

               ion:      Ho(III)
 number of charges:            8
                 g:        1.250
             units:         cm-1

 stv           2           4           6
     -0.00222222 -0.00003330 -0.00000129

 <r>k          2           4           6
      0.81000000  1.81600000  9.34500000

 shi           2           4           6
      0.53400000 -0.03060000 -0.03130000


  input coordinates

  cartesian (A)
         x            y            z            Z  
    1.3717190    0.0000000    0.5800012    0.2003000
    0.0000000    1.3717190   -0.5800012    0.2003000
   -1.3717190    0.0000000    0.5800012    0.2003000
   -0.0000000   -1.3717190   -0.5800012    0.2003000
    0.9136713    0.0635696   -1.1743841    0.2003000
   -0.0635696    0.9136713    1.1743841    0.2003000
   -0.9136713   -0.0635696   -1.1743841    0.2003000
    0.0635696   -0.9136713    1.1743841    0.2003000

  spherical (rad)
         r          theta         phi           Z  
    1.4893000    1.1707669    0.0000000    0.2003000
    1.4893000    1.9708258    1.5707963    0.2003000
    1.4893000    1.1707669    3.1415927    0.2003000
    1.4893000    1.9708258    4.7123890    0.2003000
    1.4893000    2.4792402    0.0694641    0.2003000
    1.4893000    0.6623525    1.6402604    0.2003000
    1.4893000    2.4792402    3.2110568    0.2003000
    1.4893000    0.6623525    4.7818531    0.2003000

\end{verbatim}
\end{maxipage}
\end{footnotesize}

\begin{footnotesize}
\hspace{-6.0cm}\begin{maxipage}
\begin{verbatim}

                   ROTATION                     
 _______________________________________________

 alpha =    2.0 degrees
 beta =    0.0 degrees
 gamma =    8.0 degrees

 ***********************************************
          COORDINATES AFTER ROTATION            
 ***********************************************
                                                
              Spherical coordinates             
 _______________________________________________

   #        r       theta      phi        charge
   #       [A]      [rad]     [rad]        [e]  

   1     1.48930   1.17077   0.17453     0.20030
   2     1.48930   1.97083   1.74533     0.20030
   3     1.48930   1.17077  -2.96706     0.20030
   4     1.48930   1.97083  -1.39626     0.20030
   5     1.48930   2.47924   0.24400     0.20030
   6     1.48930   0.66235   1.81479     0.20030
   7     1.48930   2.47924  -2.89760     0.20030
   8     1.48930   0.66235  -1.32680     0.20030

 ****************************************************
     Stevens Crystal Field Parameters (cm-1)         
 ****************************************************

   k   q         Akq <rk>             Bkq            
 ----------------------------------------------------

   2   0      238.51771524       -0.53003937         
   4   0      -83.48560350        0.00278007         
   4   4      643.64874438       -0.02143352         
   4  -4      592.28941789        0.01972326         
   6   0       -7.01279402        0.00000907         
   6   4      249.45424878       -0.00032272   
   6  -4      298.50706758        0.00038618         
\end{verbatim}
\end{maxipage}
\end{footnotesize}

\newpage

\begin{tiny}
\hspace{-1.0cm}\adjustbox{minipage=\textwidth,angle=-90}{%
\begin{verbatim}

Eigenvalues (cm-1) and eigenvectors

                     -8.00     -7.00     -6.00     -5.00     -4.00     -3.00     -2.00     -1.00      0.00      1.00      2.00      3.00      4.00      5.00      6.00      7.00      8.00

        0.000000  0.000000  0.786464  0.000000  0.000000 -0.000000  0.482960  0.000000  0.000000  0.000000  0.031978  0.000000  0.000000 -0.000000  0.013237  0.000000  0.000000  0.000000
                 -0.000000 -0.260893  0.000000 -0.000000 -0.000000  0.280411  0.000000  0.000000 -0.000000  0.004359 -0.000000 -0.000000  0.000000  0.017632  0.000000  0.000000  0.000000

        0.000000  0.000000 -0.000000  0.000000  0.005587  0.000000 -0.000000 -0.000000  0.027956 -0.000000 -0.000000 -0.000000  0.341071 -0.000000 -0.000000 -0.000000  0.826723 -0.000000
                 -0.000000  0.000000 -0.000000 -0.021328  0.000000  0.000000 -0.000000 -0.016127  0.000000  0.000000  0.000000 -0.442213  0.000000 -0.000000 -0.000000 -0.055861 -0.000000

       11.255613  0.000000  0.000000 -0.552390 -0.000000  0.000000  0.000000 -0.344962  0.000000  0.000000  0.000000 -0.176053 -0.000000 -0.000000  0.000000 -0.573263 -0.000000 -0.000000
                  0.000000 -0.000000  0.218205 -0.000000  0.000000 -0.000000 -0.168085  0.000000  0.000000 -0.000000  0.340964  0.000000  0.000000  0.000000  0.155300  0.000000  0.000000

       19.197764 -0.000000  0.000000 -0.570016 -0.000000 -0.000000  0.000000 -0.318614  0.000000 -0.000000 -0.000000  0.165501  0.000000  0.000000 -0.000000  0.591019  0.000000  0.000000
                 -0.000000 -0.000000  0.220359  0.000000 -0.000000 -0.000000 -0.158142 -0.000000 -0.000000  0.000000 -0.314854 -0.000000 -0.000000  0.000000 -0.155477 -0.000000  0.000000

       60.705255 -0.446141 -0.000000  0.000000  0.000000 -0.298101 -0.000000  0.000000  0.000000  0.012653  0.000000 -0.000000  0.000000  0.339785  0.000000 -0.000000  0.000000  0.448485
                 -0.045791  0.000000 -0.000000 -0.000000 -0.423514 -0.000000  0.000000 -0.000000 -0.247196  0.000000  0.000000 -0.000000 -0.390864  0.000000  0.000000  0.000000  0.000000

       71.349065 -0.485555 -0.000000  0.000000  0.000000 -0.294481 -0.000000  0.000000  0.000000 -0.000000  0.000000 -0.000000  0.000000 -0.335659  0.000000 -0.000000  0.000000 -0.488106
                 -0.049837  0.000000 -0.000000 -0.000000 -0.418371 -0.000000  0.000000  0.000000  0.000000  0.000000  0.000000  0.000000  0.386117  0.000000  0.000000  0.000000  0.000000

       77.076016 -0.000000 -0.000000 -0.000000  0.678938 -0.000000  0.000000 -0.000000  0.054157 -0.000000  0.000000  0.000000 -0.001802  0.000000 -0.000000 -0.000000  0.013560  0.000000
                  0.000000 -0.000000 -0.000000  0.538337  0.000000 -0.000000 -0.000000  0.494669  0.000000  0.000000  0.000000 -0.003302  0.000000  0.000000  0.000000 -0.037527  0.000000

       77.076016  0.000000 -0.036588  0.000000 -0.000000  0.000000 -0.003410  0.000000 -0.000000 -0.000000  0.497114  0.000000  0.000000  0.000000  0.580478  0.000000  0.000000  0.000000
                  0.000000  0.015922  0.000000  0.000000  0.000000 -0.001588 -0.000000  0.000000  0.000000  0.022548 -0.000000 -0.000000 -0.000000  0.643280  0.000000  0.000000  0.000000

      214.949594 -0.397396  0.000000 -0.000000  0.000000  0.089754 -0.000000  0.000000 -0.000000 -0.040644 -0.000000  0.000000 -0.000000 -0.102305 -0.000000  0.000000  0.000000  0.399484
                 -0.040788 -0.000000  0.000000 -0.000000  0.127515 -0.000000  0.000000  0.000000  0.794067  0.000000 -0.000000  0.000000  0.117684 -0.000000 -0.000000  0.000000  0.000000

      271.221415 -0.508944  0.000000 -0.000000  0.000000  0.280947 -0.000000  0.000000 -0.000000  0.000000 -0.000000  0.000000 -0.000000  0.320233  0.000000 -0.000000  0.000000 -0.511618
                 -0.052238 -0.000000  0.000000 -0.000000  0.399144 -0.000000  0.000000 -0.000000 -0.000000  0.000000 -0.000000  0.000000 -0.368372  0.000000  0.000000  0.000000  0.000000

      274.041991  0.000000 -0.000003 -0.000000  0.429408 -0.000000  0.000005  0.000000 -0.670210  0.000000  0.000010 -0.000000 -0.376148 -0.000000 -0.000005 -0.000000  0.158068  0.000000
                  0.000000  0.000002  0.000000 -0.130300 -0.000000  0.000001  0.000000 -0.363300 -0.000000 -0.000002 -0.000000 -0.040088  0.000000 -0.000003  0.000000  0.222221  0.000000

      274.041991 -0.000000 -0.216334  0.000000 -0.000006  0.000000  0.371345 -0.000000  0.000009 -0.000000  0.747327 -0.000000  0.000005  0.000000 -0.372245  0.000000 -0.000002 -0.000000
                 -0.000000  0.166034 -0.000000  0.000001  0.000000  0.072090 -0.000000  0.000005  0.000000 -0.150568  0.000000  0.000001 -0.000000 -0.250606  0.000000 -0.000003  0.000000

      283.277708 -0.000000 -0.000000  0.367018  0.000000  0.000000 -0.000000 -0.511431  0.000000 -0.000000 -0.000000 -0.313771 -0.000000  0.000000  0.000000  0.377214  0.000000 -0.000000
                  0.000000 -0.000000 -0.112023  0.000000  0.000000  0.000000 -0.301972 -0.000000  0.000000 -0.000000  0.504279 -0.000000 -0.000000  0.000000 -0.070433  0.000000  0.000000

      291.044315  0.000000  0.380134 -0.000000  0.000001 -0.000000 -0.727559 -0.000000 -0.000002 -0.000000  0.402545  0.000000  0.000004  0.000000 -0.183122 -0.000000 -0.000002 -0.000000
                 -0.000000 -0.304873 -0.000000 -0.000000 -0.000000 -0.125173  0.000000 -0.000001  0.000000 -0.090094  0.000000 -0.000000 -0.000000 -0.117676 -0.000000 -0.000002  0.000000

      291.044315 -0.000000  0.000002 -0.000000 -0.182568  0.000000 -0.000001  0.000000  0.083406  0.000000  0.000001  0.000000 -0.413279 -0.000000 -0.000000 -0.000000 -0.121590  0.000000
                  0.000000  0.000001 -0.000000 -0.118534  0.000000 -0.000003  0.000000  0.403984 -0.000000  0.000001 -0.000000 -0.611727  0.000000 -0.000001 -0.000000  0.471874  0.000000

      297.817674  0.371244 -0.000000 -0.000000 -0.000000 -0.262164  0.000000  0.000000 -0.000000 -0.028302 -0.000000 -0.000000 -0.000000  0.298823  0.000000  0.000000 -0.000000 -0.373195
                  0.038104  0.000000  0.000000 -0.000000 -0.372459  0.000000  0.000000  0.000000  0.552937  0.000000  0.000000 -0.000000 -0.343745  0.000000 -0.000000  0.000000  0.000000

      304.397378 -0.000000  0.000000 -0.342690  0.000000  0.000000 -0.000000  0.518335  0.000000  0.000000  0.000000 -0.335689  0.000000 -0.000000  0.000000  0.351177 -0.000000  0.000000
                  0.000000 -0.000000  0.095329  0.000000  0.000000 -0.000000  0.323736 -0.000000 -0.000000  0.000000  0.510675  0.000000  0.000000 -0.000000 -0.056553 -0.000000  0.000000

\end{verbatim}
}
\end{tiny}

\newpage

\begin{tiny}
\hspace{-0.0cm}\adjustbox{minipage=\textwidth,angle=-90}{%
\begin{verbatim}

Eigenvalues (cm-1) and eigenvectors (modulus square)

                     -8.00     -7.00     -6.00     -5.00     -4.00     -3.00     -2.00     -1.00      0.00      1.00      2.00      3.00      4.00      5.00      6.00      7.00      8.00

        0.000000  0.000000  0.686591  0.000000  0.000000  0.000000  0.311881  0.000000  0.000000  0.000000  0.001042  0.000000  0.000000  0.000000  0.000486  0.000000  0.000000  0.000000

        0.000000  0.000000  0.000000  0.000000  0.000486  0.000000  0.000000  0.000000  0.001042  0.000000  0.000000  0.000000  0.311881  0.000000  0.000000  0.000000  0.686591  0.000000

       11.255613  0.000000  0.000000  0.352749  0.000000  0.000000  0.000000  0.147251  0.000000  0.000000  0.000000  0.147251  0.000000  0.000000  0.000000  0.352749  0.000000  0.000000

       19.197764  0.000000  0.000000  0.373476  0.000000  0.000000  0.000000  0.126524  0.000000  0.000000  0.000000  0.126524  0.000000  0.000000  0.000000  0.373476  0.000000  0.000000

       60.705255  0.201139  0.000000  0.000000  0.000000  0.268229  0.000000  0.000000  0.000000  0.061266  0.000000  0.000000  0.000000  0.268229  0.000000  0.000000  0.000000  0.201139

       71.349065  0.238247  0.000000  0.000000  0.000000  0.261753  0.000000  0.000000  0.000000  0.000000  0.000000  0.000000  0.000000  0.261753  0.000000  0.000000  0.000000  0.238247

       77.076016  0.000000  0.000000  0.000000  0.750763  0.000000  0.000000  0.000000  0.247631  0.000000  0.000000  0.000000  0.000014  0.000000  0.000000  0.000000  0.001592  0.000000

       77.076016  0.000000  0.001592  0.000000  0.000000  0.000000  0.000014  0.000000  0.000000  0.000000  0.247631  0.000000  0.000000  0.000000  0.750763  0.000000  0.000000  0.000000

      214.949594  0.159587  0.000000  0.000000  0.000000  0.024316  0.000000  0.000000  0.000000  0.632194  0.000000  0.000000  0.000000  0.024316  0.000000  0.000000  0.000000  0.159587

      271.221415  0.261753  0.000000  0.000000  0.000000  0.238247  0.000000  0.000000  0.000000  0.000000  0.000000  0.000000  0.000000  0.238247  0.000000  0.000000  0.000000  0.261753

      274.041991  0.000000  0.000000  0.000000  0.201370  0.000000  0.000000  0.000000  0.581169  0.000000  0.000000  0.000000  0.143094  0.000000  0.000000  0.000000  0.074368  0.000000

      274.041991  0.000000  0.074368  0.000000  0.000000  0.000000  0.143094  0.000000  0.000000  0.000000  0.581169  0.000000  0.000000  0.000000  0.201370  0.000000  0.000000  0.000000

      283.277708  0.000000  0.000000  0.147251  0.000000  0.000000  0.000000  0.352749  0.000000  0.000000  0.000000  0.352749  0.000000  0.000000  0.000000  0.147251  0.000000  0.000000

      291.044315  0.000000  0.237449  0.000000  0.000000  0.000000  0.545010  0.000000  0.000000  0.000000  0.170159  0.000000  0.000000  0.000000  0.047381  0.000000  0.000000  0.000000

      291.044315  0.000000  0.000000  0.000000  0.047381  0.000000  0.000000  0.000000  0.170159  0.000000  0.000000  0.000000  0.545010  0.000000  0.000000  0.000000  0.237449  0.000000

      297.817674  0.139274  0.000000  0.000000  0.000000  0.207456  0.000000  0.000000  0.000000  0.306540  0.000000  0.000000  0.000000  0.207456  0.000000  0.000000  0.000000  0.139274

      304.397378  0.000000  0.000000  0.126524  0.000000  0.000000  0.000000  0.373476  0.000000  0.000000  0.000000  0.373476  0.000000  0.000000  0.000000  0.126524  0.000000  0.000000

\end{verbatim}
}
\end{tiny}

\newpage

\subsection*{sus.out}

\begin{small}
\begin{maxipage}
\begin{verbatim}
******************************************************
**  output file generated by SIMPRE version 1.1  **
******************************************************

 ____________________________________________________

 .######..####.##.....##.########..########..########
 ##....##..##..###...###.##.....##.##.....##.##......
 ##........##..####.####.##.....##.##.....##.##......
 .######...##..##.###.##.########..########..######..
 ......##..##..##.....##.##........##...##...##......
 ##....##..##..##.....##.##........##....##..##......
 .######..####.##.....##.##........##.....##.########
 ____________________________________________________

 ****************************************************

Magnetic susceptibility from     2.000 K to   300.000 K , number of points:  25,
at a magnetic field equal to  0.100 T

       T (K)       XzT (emu·K/mol)   XxT (emu·K/mol)   XyT (emu·K/mol)  XavT (emu·K/mol)    

        2.00         18.955733          1.316741          1.316741          7.196405
        2.46         19.076473          1.620717          1.620717          7.439302
        3.04         19.144374          1.990367          1.990367          7.708369
        3.74         19.161943          2.432303          2.432303          8.008849
        4.61         19.128331          2.946601          2.946601          8.340511
        5.68         19.045000          3.523915          3.523915          8.697610
        7.00         18.918390          4.145715          4.145715          9.069940
        8.62         18.759376          4.789199          4.789199          9.445925
       10.63         18.581103          5.435347          5.435347          9.817266
       13.09         18.396426          6.075424          6.075424         10.182424
       16.13         18.215369          6.711116          6.711116         10.545867
       19.88         18.043345          7.347429          7.347430         10.912735
       24.49         17.881396          7.982776          7.982776         11.282316
       30.18         17.729112          8.603590          8.603590         11.645430
       37.19         17.588419          9.188248          9.188248         11.988305
       45.82         17.464479          9.717944          9.717944         12.300122
       56.46         17.361256         10.186590         10.186590         12.578146
       69.57         17.274018         10.603268         10.603268         12.826851
       85.72         17.185100         10.986455         10.986455         13.052670
      105.63         17.068476         11.353900         11.353900         13.258759
      130.15         16.902061         11.713934         11.713934         13.443310
      160.37         16.679318         12.063114         12.063114         13.601848
      197.60         16.411622         12.390705         12.390705         13.731011
      243.47         16.121081         12.685677         12.685677         13.830811
      300.00         15.830666         12.941380         12.941380         13.904476
\end{verbatim}
\end{maxipage}
\end{small}

\subsection*{mag.out}

\begin{small}
\begin{maxipage}
\begin{verbatim}
******************************************************
**  output file generated by SIMPRE version 1.1  **
******************************************************

 ____________________________________________________

 .######..####.##.....##.########..########..########
 ##....##..##..###...###.##.....##.##.....##.##......
 ##........##..####.####.##.....##.##.....##.##......
 .######...##..##.###.##.########..########..######..
 ......##..##..##.....##.##........##...##...##......
 ##....##..##..##.....##.##........##....##..##......
 .######..####.##.....##.##........##.....##.########
 ____________________________________________________

 ****************************************************

Magnetization at     2.000 K until     5.000 T with a step of:     0.250 T,
with an integration angle of    1.4754°

       H (T)           Mz (bm)           Mx (bm)           My (bm)           Mav(bm) 

      0.0000          0.008648          0.000590          0.000590          0.002986
      0.2500          3.874290          0.294986          0.294986          1.413101
      0.5000          6.001448          0.587175          0.587175          2.465570
      0.7500          6.811746          0.875030          0.875030          3.157458
      1.0000          7.077595          1.156548          1.156548          3.623112
      1.2500          7.162470          1.430015          1.430015          3.963514
      1.5000          7.191563          1.693611          1.693611          4.233418
      1.7500          7.203675          1.946278          1.946278          4.460310
      2.0000          7.210334          2.186941          2.186941          4.658530
      2.2500          7.215649          2.415005          2.415005          4.835540
      2.5000          7.220477          2.629926          2.629926          4.995930
      2.7500          7.226344          2.832038          2.832038          5.142631
      3.0000          7.230727          3.021478          3.021478          5.277348
      3.2500          7.233305          3.198434          3.198434          5.401804
      3.5000          7.237836          3.363568          3.363568          5.516760
      3.7500          7.242344          3.518224          3.518224          5.623717
      4.0000          7.248820          3.661059          3.661059          5.723251
      4.2500          7.253710          3.795067          3.795067          5.816167
      4.5000          7.255739          3.920576          3.920576          5.902834
      4.7500          7.260164          4.037972          4.037972          5.984287
      5.0000          7.266569          4.146565          4.146565          6.060429
\end{verbatim}
\end{maxipage}
\end{small}

\newpage
\section{Example 3: Diphenylbis(pyrazolborate) complex of a user-defined atom}

This example illustrates the use of the {\em user-defined} ion. Notice the
three mandatory lines after the coordinate list.

\subsection*{simpre.par}

\begin{small}
\begin{maxipage}
\begin{verbatim}
      PARAMETER (idtot= 17)! dimension of full energy matrix
c
      PARAMETER (iuni=1)  ! units (1=K, 2=meV, 3=cm-1)
c
      PARAMETER (icoo=2)  ! spherical coordinates=1, cartesian coordinates=2
c
      PARAMETER (ishi=1)  ! use of Sternheimer shielding parameters [1(yes)/0(no)]
c
      PARAMETER (irot=0)  ! rotation for a compact ground-state wave function [1(yes)/0(no)]
c
      PARAMETER (isus=0)  ! magnetic susceptibility calculation [1(yes)/0(no)]
c
      PARAMETER (imag=0)  ! magnetization calculation [1(yes)/0(no)]
c
      PARAMETER (ieig=2)  ! write eigenvalues 
c                                   [no(0)/yes(1)/yes+eigenvectors(2)]
c
c------------------------------------------------------------------------------
c* The following parameter is defined to minimize underflow problems
        PARAMETER (EPS    = 1.D-12)
c
        PARAMETER (ICHAR =  12)! number of charges
\end{verbatim}
\end{maxipage}
\end{small}

\subsection*{simpre.dat}

\begin{small}
\begin{maxipage}
\begin{verbatim}
1=Ce3+, 2=Pr3+, 3=Nd3+, 4=Pm3+, 5=Sm3+, 6=Tb3+, 7=Dy3+, 8=Ho3+, 9=Er3+, 10=Tm3+, 11=Yb3+, 12=user

 12    !ion code (from 1 to 12, see above)
  6    !number of effective charges
  1   -0.8379900    0.5157900   -0.7168800    0.0246667
  2   -0.8881000    0.4130100    0.7456400    0.0246667
  3    0.8931500    0.4530000    0.6378900    0.0246667
  4    0.8289000    0.4798100   -0.8321300    0.0246667
  5   -0.0011300   -0.9885800   -0.6730100    0.0246667
  6   -0.0022500   -0.9781800    0.7919800    0.0246667



           X(III)                9    0.71757575758   !SYMBOL, 2*J, GJ 
   -0.00642791551   -0.00029110772   -0.00003798795   !2nd,4th and 6th order Stevens parameters
           0.8300           0.0026          -0.0390   !2nd,4th and 6th order shielding parameters
           2.3460          10.9060          90.5440   !2nd,4th and 6th order expected values of radius	
\end{verbatim}
\end{maxipage}
\end{small}

\subsection*{simpre.out}

\begin{scriptsize}
\hspace{-6.0cm}\begin{maxipage}
\begin{verbatim}
******************************************************
**  output file generated by SIMPRE version 1.1  **
******************************************************
 ____________________________________________________

 .######..####.##.....##.########..########..########
 ##....##..##..###...###.##.....##.##.....##.##......
 ##........##..####.####.##.....##.##.....##.##......
 .######...##..##.###.##.########..########..######..
 ......##..##..##.....##.##........##...##...##......
 ##....##..##..##.....##.##........##....##..##......
 .######..####.##.....##.##........##.....##.########
 ____________________________________________________

 ****************************************************

               ion:       X(III)
 number of charges:            6
                 g:        0.718
             units:          K  

 stv           2           4           6
     -0.00642792 -0.00029111 -0.00003799

 <r>k          2           4           6
      2.34600000 10.90600000 90.54400000

 shi           2           4           6
      0.83000000  0.00260000 -0.03900000

  input coordinates

  cartesian (A)
         x            y            z            Z  
   -0.8379900    0.5157900   -0.7168800    0.0246667
   -0.8881000    0.4130100    0.7456400    0.0246667
    0.8931500    0.4530000    0.6378900    0.0246667
    0.8289000    0.4798100   -0.8321300    0.0246667
   -0.0011300   -0.9885800   -0.6730100    0.0246667
   -0.0022500   -0.9781800    0.7919800    0.0246667

  spherical (rad)
         r          theta         phi           Z  
    1.2174496    2.2004162    2.5898478    0.0246667
    1.2309662    0.9201064    2.7062947    0.0246667
    1.1873624    1.0036436    0.4693859    0.0246667
    1.2687526    2.2861238    0.5247240    0.0246667
    1.1959240    2.1685090   -1.5719394    0.0246667
    1.2585998    0.8902023   -1.5730965    0.0246667

 ****************************************************
     Stevens Crystal Field Parameters ( K  )
 ****************************************************

   k   q         Akq <rk>             Bkq            
 ----------------------------------------------------

   2   0       23.41065657       -0.15048172   
   2   1      -29.92791591        0.19237411   
   2  -1        2.24600853        0.01443715   
   2   2       23.94518427       -0.15391762   
   2  -2        1.50712273        0.00968766   
   4   0     -376.94384308        0.10973126   
   4   1      416.12483946       -0.12113715   
   4  -1      389.14373223        0.11328274   
   4   2       65.88635646       -0.01918003   
   4  -2      -29.77024960        0.00866635   
   4   3      237.84649020       -0.06923895   
   4  -3      -14.24150337        0.00414581   
   4   4      140.39121655       -0.04086897   
   4  -4       37.87978345        0.01102710   
   6   0      142.22043705       -0.00540266   
   6   1     -135.05761204        0.00513056   
   6  -1     -146.93808243        0.00558188   
   6   2      -61.68145480        0.00234315   
   6  -2       15.13192797        0.00057483   
   6   3      148.45859672       -0.00563964   
   6  -3      222.63072628        0.00845728   
   6   4      278.09980792       -0.01056444   
   6  -4      -30.03866824        0.00114111   
   6   5     2018.60095274       -0.07668251   
   6  -5    -3618.44512037        0.13745731   
   6   6    -1510.41687012        0.05737764   
   6  -6       67.05066463        0.00254712   
\end{verbatim}
\end{maxipage}
\end{scriptsize}

\newpage

\begin{scriptsize}
\hspace{-6.0cm}\begin{maxipage}
\begin{verbatim}
Eigenvalues ( K  ) and eigenvectors

                     -4.50     -3.50     -2.50     -1.50     -0.50      0.50      1.50      2.50      3.50      4.50

        0.000000 -0.022218  0.077861 -0.400821 -0.065981 -0.003519  0.014811  0.053240 -0.045523  0.196565  0.014089
                 -0.033810 -0.018635 -0.795638  0.061204  0.001481 -0.021880  0.115633  0.008663  0.363670 -0.000000

        0.000000 -0.007738  0.411870  0.017760  0.125874  0.010152 -0.000695 -0.014913 -0.885041 -0.027186 -0.040457
                 -0.011774 -0.035449  0.042801 -0.019010 -0.024394 -0.003754  0.088753  0.101978 -0.075303  0.000000

      279.145959  0.196216  0.175393 -0.087109 -0.558219  0.030353  0.090522 -0.275267  0.055419 -0.056453  0.249567
                 -0.363215 -0.023435  0.200441 -0.007118 -0.006526  0.058910  0.524343  0.003256  0.083364  0.000000

      279.145959  0.118619  0.100178  0.023476  0.592164 -0.008806 -0.020169 -0.259059  0.217756  0.103983 -0.412827
                 -0.219575 -0.010046 -0.050307  0.007034 -0.107644  0.023603  0.494518  0.018629 -0.143176  0.000000

      582.517724 -0.183831 -0.188890 -0.034632  0.048243 -0.063245  0.422689 -0.065625 -0.090902  0.003939 -0.024206
                 -0.685349  0.249931 -0.063749 -0.041454 -0.158023  0.144996 -0.349640  0.131904 -0.087000  0.000000

      582.517724  0.006271 -0.083009 -0.103850 -0.354704 -0.249553 -0.169013  0.027540 -0.070544 -0.192462 -0.709576
                  0.023379  0.026343  0.121971  0.027198 -0.370693 -0.020147 -0.057336 -0.016934  0.247191  0.000000

      858.416889  0.047289  0.653795  0.028543 -0.032383 -0.038308  0.276854  0.008060  0.265042 -0.000872 -0.110868
                 -0.041785 -0.462127 -0.008626  0.023283 -0.050885  0.159595 -0.344055 -0.212647 -0.012469  0.000000

      858.416889 -0.083081 -0.007603  0.339419 -0.233856  0.101791 -0.004987 -0.039683 -0.027101  0.795933 -0.063105
                  0.073411 -0.009921 -0.016146 -0.252488 -0.302915 -0.063497  0.003995  0.012436 -0.086605  0.000000

     1114.397935 -0.292810  0.083644  0.016138 -0.067247  0.166577 -0.748419 -0.153497  0.023531  0.022763 -0.041277
                 -0.397060 -0.142917 -0.055444 -0.024945  0.233183  0.097817 -0.184369 -0.046268 -0.080085  0.000000

     1114.397935  0.024498 -0.050945  0.023272 -0.239488  0.365471  0.286537  0.059989 -0.035045  0.065379 -0.493350
                  0.033220  0.065852 -0.046399 -0.014112  0.660402 -0.004332  0.039317  0.045895 -0.152142  0.000000


Eigenvalues ( K  ) and eigenvectors (modulus square)

                     -4.50     -3.50     -2.50     -1.50     -0.50      0.50      1.50      2.50      3.50      4.50

        0.000000  0.001637  0.006410  0.793697  0.008099  0.000015  0.000698  0.016206  0.002147  0.170894  0.000199

        0.000000  0.000199  0.170894  0.002147  0.016206  0.000698  0.000015  0.008099  0.793697  0.006410  0.001637

      279.145959  0.170426  0.031312  0.047765  0.311659  0.000964  0.011665  0.350707  0.003082  0.010137  0.062284

      279.145959  0.062284  0.010137  0.003082  0.350707  0.011665  0.000964  0.311659  0.047765  0.031312  0.170426

      582.517724  0.503497  0.098145  0.005263  0.004046  0.028971  0.199690  0.126555  0.025662  0.007585  0.000586

      582.517724  0.000586  0.007585  0.025662  0.126555  0.199690  0.028971  0.004046  0.005263  0.098145  0.503497

      858.416889  0.003982  0.641009  0.000889  0.001591  0.004057  0.102119  0.118439  0.115466  0.000156  0.012292

      858.416889  0.012292  0.000156  0.115466  0.118439  0.102119  0.004057  0.001591  0.000889  0.641009  0.003982

     1114.397935  0.243394  0.027422  0.003335  0.005144  0.082122  0.569699  0.057553  0.002694  0.006932  0.001704

     1114.397935  0.001704  0.006932  0.002694  0.057553  0.569699  0.082122  0.005144  0.003335  0.027422  0.243394

\end{verbatim}
\end{maxipage}
\end{scriptsize}

\chapter{A step beyond}
\label{community}

\section{Different electrostatic models}
\label{models}

In the past century, many electrostatic models have been proposed to deal with
the problem of estimating the Crystal Field parameters of lanthanoid ions,
briefly reviewed by P. Porcher.$^1$ Being fundamentally similar, most of them are
compatible with \simpre, either in its standard state or after small
modifications (see below). We discuss here the differences between the most
important of these models.

\begin{itemize}
\item the Point Charge Electrostatic Model (PCEM)$^2$ is the simplest possibility:
substitute each atom in the first coordination sphere for point charges by
values determined by the valence of each atom. It is a rough approximation for
ionic systems such as Ln:LiYF$_4$ but fails in cases of less ionic bonding. To
implement the classical PCEM in \simpre, the user just needs to set the control
variable {\verb ishi = 0} to cancel Sternheimer's shielding parameters, while
{\verb ishi = 1} includes this shielding which decreases the second order CF
parameters, which generally constitutes an improvement in the quality of the
results.
\item The Effective Charge Model (ECM)$^3$ allows reproducing the covalent effects by
changing both the position and the magnitude of the charges, thus being similar
to the REC model (see below). It also includes as effective charges all atoms
in a 100 \AA{} radius sphere. It is able to reproduce the CF parameters in many
cases, being a more limited approximation in systems with a higher covalent
character. ECM can be implemented by means of the user-defined atom, where the
$\langle r^k\rangle$ and $\sigma_k$ should be substituted accordingly to
include the antishielding effect, see Table 17 in ref.$^4$.  
\item The Simple Overlap Model (SOM)$^5$ is more versatile than ECM and as a
major advantage, does only need to consider the first coordination sphere. The
effective charges are very similar to those in ECM, but SOM does not include
free parameters. SOM can be implemented in \simpre with {\verb ishi = 0} and
calculating charge and distance according to the following formulas (for
details see ref. [6]): the input distance $r_i$ is related to the
crystallographic distance $R_i$ by $r_i=R_i(1\pm\rho_i)/2$ with
$\rho_i=0.05\cdot(R_0/R_\mu)^{3.5}$. The input charge $Z_i$ is related to a
tabulated charge $g_i$ by $Z_i=|g_i|\cdot\rho_i$ 
\item The Angular Overlap Model (AOM)$^{6}$ includes specific parameters to
discriminate the angular part of the crystal field effect. Like SOM (and REC
and LPEC), it allows to classify the ligands in terms of their relative
covalence, and allows an interpretation of the CF parameters in terms of a
connection with the nature of the ligands and their coordinates.  Note that in
all previous to AOM the ration between the CF parameters found by the PCEM does
not change.  In this aspect, AOM is similar to the LPEC model (see below).
Implementing AOM in \simpre would require more elaborate changes.
\item The Radial Effective Charge (REC) model the Lone Pair Effective Charge
(LPEC) model are our basic tools and thus are explained with more detail beyond.
\end{itemize}

All these models are relatively easy to apply, as they only require the
crystallographic positions. They are also oversimplified when compared with the
whole list of interactions that one could take into account. However, they
produce a reasonable order of magnitude of the CF parameters. This is of
crucial importance, if we keep in mind the great number of phenomenological
parameters that have to be varied for the simulation of a $4f_N$ configuration
energy level scheme. Thus, these starting values are as essential for the
spectroscopic simulation as atomic positions are for structure
determinations.

{\bf References}

{\rm [1]} P. Porcher, M. C. Dos Santos, O. Malta, {\emph Phys. Chem. Chem.
Phys.}, 1999, {\bf 1}, 397

{\rm [2]} H. Bethe, {\emph Ann. Phys.} 1929, {\bf 3}, 133

{\rm [3]} see e.g. B. R. Judd, {\emph Operators Techniques in Atomic
Spectroscopy}, McGraw-Hill, 1963

{\rm [4]} C. A. Morrison, {\emph Lectures on Crystal Field Theory}, HDL-SR-82-2,
Harry Diamond Laboratories Report, 1982

{\rm [5]} O. L. Malta, {\emph Chem. Phys. Lett.}, 1982, {\bf 82}, 27;
1982, {\bf 88}, 353 

{\rm [6]} C. K. J\o rgensen, R. Pappalardo, H. B. Schmidtke, {\emph J. Chem.
Phys.}, 1963, {\bf 39}, 1422

\section{Effective Charge Models: REC and LPEC}

The rationalizing and predictive capacity of \simpre is strongly enhanced by a wise
use of effective charges that are displaced from the crystallographic position
of the nuclei. Indeed, extensive testing has demonstrated that na\"ive use of
formal charges and crystallographic coordinates rarely results in an accurate
prediction of the magnetic properties due to the intrinsic limitations of the
Point Charge Electrostatic Model (PCEM). If, instead, the value and displacements
of the charges are allowed to be fitted as free parameters to some property,
the description of the system is often realistic enough to predict other
properties which have not been used in the fit. Likewise, the description of
the ligands can be such that the properties of analogous compounds can be
predicted. It is thus strongly recommended to use either the Radial Effective
Charge (REC) or, when needed, the Lone Pair Effective Charge (LPEC) models,
for a more realistic description of the interaction between the lanthanoid and
the donor atoms. When high-quality spectroscopic information is available, the
obtained crystal field parameters can be used as a starting point for fitting
purposes.

The use of {\bf REC} is very simple: for a complex that is coordinated by a given
ligand, such as fluorine or a pyrazol-kind nitrogen (see examples 2 and 3), a
set of calculations is launched with two free parameters: the effective charge
and a reduction in the effective radius. The preparation of the input and the
comparison of the outputs with the experimental data can be easily done by
hand, or with the aid of an external program, or an improved version of \simpre
with extra loops. As usually different solutions are possible, it is best to do
either a collective fit to several isostructural complexes of different
lanthanoids, or use spectroscopic data instead of a $\chi T$ curve, or both. 

Of course, this strategy is time-consuming and can even be impractical when
dealing with heteroleptic complexes, where the number of free parameters is
higher. Fortunately, there is a growing body of parameterized ligands, with the
final goal of building a {\bf general library}, as explained below. In simple
cases (homoleptic or with abundance of experimental information), reusing
published parameters can serve to check their validity, and in complex cases
(heteroleptic and/or limited experimental information) it is arguably the best
way to reduce or eliminate the overparameterization problem.

The use of {\bf LPEC} is very similar to the use of REC, but additionally it
requires the determination of a unit vector pointing in the direction of the
lone pair, usually using chemical arguments and simple geometry. In the case of
the phthalocyaninato nitrogens, this can be done by taking into account the
planarity of the ligand and the $sp^2$ character of the lone pair, which
displays angles close to 120$^\circ$ with the other two $sp^2$ orbitals that
form the N-C bonds. In a way, it is reminiscent of the Angular Overlap Model -a
more chemical approach- which discriminates the angular part of the crystal
field (see section~\ref{models}).

\subsection{Library of ligands}

The effective point charge model used by \simpre can be seen as part of a
rediscovery, in a new context, of the Crystal Field Theory of the 1960s by the
molecular magnetism community. However, what made tools like the Angular
Overlap Model powerful was the availability of consistent parameters describing
the influence of a given ligand on the electronic structure of the central ion.
This was achieved by measuring the optical properties of large series of
complexes, and distilling optimal parameter values from the observed level
splittings. Thus, we need a similar work to be done with \simpre, so that a
{\bf general library of ligands} can be established which parameterizes the
most common ligands in terms of the Radial Effective Charge (REC) or Lone Pair
Effective Charge (LPEC) models. This work has already started, and the results
are very encouraging for a simple and inexpensive estimate of the magnetic
properties and ground state wave-function. Of course, the availability of
sophisticated experimental information is needed a proper understanding of the
full set of energy levels and wavefunctions.

At the time of this writing, the provisional parameterization of following
ligands has been published ($D_h, D_r$ in $\AA$, $Z$ in electron charges, note
that the influence of the different Coordination Index (CI) has not yet been
established):
\marginpar{Save time and improve your results building upon previous knowledge!}
\begin{itemize}
\item fluoride, CI = 6: $D_r \simeq 1.03$, $Z \simeq 0.20$
\item chloride, CI = 6: $D_r \simeq 1.00$, $Z \simeq 0.26$
\item bromide, CI = 6: $D_r \simeq 0.90$, $Z \simeq 0.45$
\item pyrazolyl nitrogens, CI = 6-9: $1.3 \leq D_r \leq 1.5$, $0.015 \leq Z \leq 0.03$
\end{itemize}

Moreover, the following ligands are currently under study and their provisional
parameters might be available by request:
\begin{itemize}
\item aromatic carbon ring
\item vacant polyoxotungstate 
\item amine oxide
\item (dialkylphosphito-P)cobaltate
\end{itemize}

\section{Community}

The ambitious goal of the general library of ligands can only be achieved by
means of a community work, where different groups extract these parameters from
the highest quality data available and then check their validity in analogous
compounds.  An email list is maintained by the authors to facilitate the
communication within the user community. 
\marginpar{Sign up for the email list: be up-to-date and increase the
visibility of your latest results!} 
This allows users to share their (published) results and be updated on
both experimental data and theoretical advances. These progressively build up a
general library of ligands, thus making \simpre an increasingly powerful tool.
This email list is also used to keep the community up-to-date on new versions
and patches of the program as soon as they are available.  

\section{Extensions and patches}

As the source code of \simpre is distributed, it is possible for users to adapt
it to their own needs. Additionally, the authors are still working on their
own improvements and will distribute extended or patched codes by request, as
soon as they are usable.  Among the problems that may be solved with extra
subroutines or small modifications that are currently work-in-progress are:
\begin{itemize} \item automatic determination of the easy axis of magnetization
\item obtaining the angular dependence of the magnetic susceptibility \item
considering the hyperfine coupling to the nuclear spin
\item automatic fitting of the REC or LPEC parameters to available $\chi T$ vs
$T$ or spectroscopic data
\item plotting the magnetic field dependence of the energy levels
\end{itemize}

\newpage

\section{Bibliography}

The following works rely on \simpre and contain results that can be useful for
the \simpre community:

\begin{tabular}{c|c}
Reference & Summary \\
\hline
\hline
J. Comput. Chem. 2013,& Program is presented.                 \\
34, 1961              &                                      \\
\hline
Inorg. Chem. 2012,    & Theoretical background is given. \\
51, 12565             & \simpre is applied for magnetostructural studies. \\
\hline
Dalton Trans., 2012,  & REC and LPEC models are introduced. \\
41, 13705             & Parameterization of fluoride, chloride,  \\
                      & bromide, phthalocyaninato ligands. \\
\hline
Chem. Sci., 2013,     & Uranium is introduced as user-defined metal. \\
4, 938                & Limitations for actinides are discussed. \\
                      & Parameterization of pyrazole ligands. \\
\hline
Polyhedron, 2013,     & Library of ligands successfully applied,  \\
66, 39                & pyrazole ligand parameters confirmed.      \\
\hline

\end{tabular}

\chapter{Appendix: Operators}

Explicit list of the extended Stevens operators $O_k^q (J)$ used in the \simpre code. 

\begin{maxipage}
$X=J\cdot(J+1)$

\begin{tabular}{|lll|}
\hline
&&\\
$O_2^0        $&          &$= \left[3J_z^2-X\right]$ \\
&&\\
\hline
&&\\
$O_2^1  $&$\equiv O_2^1(c) $&$= 1/4    \left[ J_z\JaaJe + \JaaJe J_z \right] $ \\
&&\\
$O_2^{-1} $&$\equiv O_2^1(s) $&$= -i/4 \left[ J_z\JaeJe + \JaeJe J_z \right] $ \\
&&\\
$O_2^2  $&$\equiv O_2^2(c) $&$= 1/2    \left[ \JaaJed \right] $ \\
&&\\
$O_2^{-2} $&$\equiv O_2^2(s) $&$= -i/2 \left[ \JaeJed \right] $ \\
&&\\
\hline
\hline
&&\\
$O_4^0        $&              &$=      \left[35J_z^4-\left(30X-25\right)J_z^2+3X^2-6X\right]$ \\
&&\\
\hline
&&\\
$O_4^1  $&$\equiv O_4^1(c)   $&$= 1/4  \left[ \JaaJe\left(7J_z^3-(3X+1)J_z\right)+\left(7J_z^3-(3X+1)J_z\right)\JaaJe  \right] $ \\
&&\\
$O_4^{-1} $&$\equiv O_4^1(s) $&$= -i/4 \left[ \JaeJe\left(7J_z^3-(3X+1)J_z\right)+\left(7J_z^3-(3X+1)J_z\right)\JaeJe  \right] $ \\
&&\\
$O_4^2  $&$\equiv O_4^2(c)   $&$= 1/4  \left[ \JaaJed\left(7J_z^2-X-5\right)+ \left(7J_z^2-X-5\right)\JaaJed \right] $\\
&&\\
$O_4^{-2} $&$\equiv O_4^2(s) $&$= -i/4 \left[ \JaeJed\left(7J_z^2-X-5\right)+ \left(7J_z^2-X-5\right)\JaeJed \right] $\\
&&\\
$O_4^3   $&$\equiv O_4^3(c)$&$= 1/4 \left[ \JaaJet J_z + J_z\JaaJet \right]$ \\
&&\\
$O_4^{-3}$&$\equiv O_4^3(s)$&$=-i/4 \left[ \JaeJet J_z + J_z\JaeJet \right]$ \\
&&\\
$O_4^4   $&$\equiv O_4^4(c)$&$= 1/2 \left[ \JaaJec + \JaaJec \right]$ \\
&&\\
$O_4^{-4}$&$\equiv O_4^4(s)$&$=-i/2 \left[ \JaeJec + \JaeJec \right]$ \\
&&\\
\hline
\end{tabular}

\end{maxipage}
\pagebreak
\begin{maxipage}

$X=J\cdot(J+1)$

\begin{tabular}{|lll|}
\hline
&&\\
$O_6^0        $&              &$=      [231J_z^6-\left(315X-735\right)J_z^4+$\\
&&$\left(105X^2-525X+294\right)J_z^2 -5X^3 +40X^2 -60X ]$ \\
&&\\
\hline
&&\\
$O_6^1    $&$\equiv O_6^1(c) $&$= 1/4  [ \JaaJe\left(33J_z^5-(30X-15)J_z^3+(5X^2-10X+12)J_z\right)$\\
&&$+\left(33J_z^5-(30X-15)J_z^3+(5X^2-10X+12)J_z\right)\JaaJe ]$ \\
&&\\
$O_6^{-1} $&$\equiv O_6^1(s) $&$= -i/4 [ \JaeJe\left(33J_z^5-(30X-15)J_z^3+(5X^2-10X+12)J_z\right)$\\
&&$+\left(33J_z^5-(30X-15)J_z^3+(5X^2-10X+12)J_z\right)\JaaJe ]$ \\
&&\\
$O_6^2    $&$\equiv O_6^2(c) $&$= 1/4  [ \JaaJed\left(33J_z^4-(18X+123)J_z^2+X^2+10X+102\right)$\\
&&$+\left(33J_z^4-(18X+123)J_z^2+X^2+10X+102\right)\JaaJed ]$ \\
&&\\
$O_6^{-2} $&$\equiv O_6^2(s) $&$= -i/4 [ \JaeJed\left(33J_z^4-(18X+123)J_z^2+X^2+10X+102\right)$\\
&&$+\left(33J_z^4-(18X+123)J_z^2+X^2+10X+102\right)\JaeJed ]$ \\
&&\\
$O_6^3    $&$\equiv O_6^3(c) $&$= 1/4  \left[ \JaaJet\left(11J_z^3-(3X+59)J_z\right)+\left(11J_z^3-(3X+59)J_z\right)\JaaJet \right]$ \\
&&\\
$O_6^{-3} $&$\equiv O_6^3(s) $&$= -i/4 \left[ \JaeJet\left(11J_z^3-(3X+59)J_z\right)+\left(11J_z^3-(3X+59)J_z\right)\JaeJet \right]$ \\
&&\\
$O_6^4    $&$\equiv O_6^4(c) $&$= 1/4  \left[ \JaaJec\left(11J_z^2-X-38\right)+\left(11J_z^2-X-38\right)\JaaJec \right]$ \\
&&\\
$O_6^{-4} $&$\equiv O_6^4(s) $&$= -i/4 \left[ \JaeJec\left(11J_z^2-X-38\right)+\left(11J_z^2-X-38\right)\JaeJec \right]$ \\
&&\\
$O_6^5    $&$\equiv O_6^5(c) $&$= 1/4  \left[ \JaaJeq J_z + J_z \JaaJeq \right]$ \\
&&\\
$O_6^{-5} $&$\equiv O_6^5(s) $&$= -i/4 \left[ \JaeJeq J_z + J_z \JaeJeq \right]$ \\
&&\\
$O_6^6    $&$\equiv O_6^6(c) $&$= 1/2  \left[ \JaaJes \right]$ \\
&&\\
$O_6^{-6} $&$\equiv O_6^6(s) $&$= -i/2 \left[ \JaeJes \right]$ \\
&&\\
\hline
\end{tabular}

\end{maxipage}
\end{document}